\documentclass[letter,a4,traditabstract]{aa}
\usepackage{natbib} 
\usepackage{graphicx}
\usepackage{txfonts}

\begin{document}

\title{Activity restart -- a key to explaining the morphology of J1211+743}

\author{Andrzej Marecki}

\offprints{Andrzej Marecki \email{amr@astro.uni.torun.pl}}

\institute{Toru\'n Centre for Astronomy, N. Copernicus University,
           87-100 Toru\'n, Poland}

\date{Received 21 May 2012 / Accepted 15 June 2012}

\abstract{J1211+743 is a giant radio galaxy with a one-sided jet and two 
asymmetric lobes, one of which is of Fanaroff-Riley (FR) type\,II with a 
hotspot and the other is a diffuse relic devoid of a hotspot. The jet points 
towards the latter lobe, which is difficult to explain in a standard way 
within the double-lobed radio source paradigm. Here, I~propose to 
assume that the nucleus of J1211+743 has undergone a re-ignition of 
activity and its lobes, presumably both originally of FR\,II type, 
represent an earlier active phase, while the jet represents the current one. 
The asymmetry of the lobes is a consequence of the orientation of the source 
combined with an activity switch-off that occurred between two active 
periods. The relic lobe is on the near side with regard to the observer, 
whereas the radiation from the far-side lobe arrives significantly later 
owing to its longer distance to the observer. The far-side lobe is 
thus perceived to have not yet decayed. On the other hand, the jet behaves 
in a standard way, i.e. its projected orientation reflects the near side of 
the source. Hence, we are able to explain why the location of the relic lobe 
correlates with the direction of the jet.}


\maketitle


\section{Introduction}

\object{J1211+743} (4CT\,74.17.01) is a $z=0.107$ radio galaxy belonging to 
the Abell\,1500 cluster. An early multi-frequency radio survey of J1211+743 
was conducted by \citet{vBW1981} with the Westerbork Synthesis Radio 
Telescope (WSRT) at 609, 1415, and 4996\,MHz. \citet{Zhao89} and 
\citet{Lara2001} presented good-quality images resulting from the Very Large 
Array (VLA) observations at 1.4\,GHz. The most recent maps of J1211+743 were 
published by \citet{PNSS} (hereafter PNSS) following the observations at 
239, 333, and 607\,MHz carried out using the Giant Metrewave Radio Telescope 
(GMRT) and at 4885\,MHz using the VLA. The 607-MHz GMRT image shows the 
source with the greatest detail available to date. On the basis of the above 
radio imagery, it can be firmly asserted that the radio structure of 
J1211+743 consists of a one-sided jet originating in the infra\-red object 
\object{2MASX\,J12115859+7419038} and two lobes that straddle this object. 
The projected linear size of the whole radio structure is 846\,kpc (PNSS).

J1211+743 has two interesting properties: its lobe asymmetry and 
``non-standard'' orientation of the jet. While the southern lobe is clearly 
of Fanaroff-Riley (FR) type\,II \citep{FR74}, the northern one is devoid of 
a hotspot and has a steeper spectrum ($\alpha=-1.02$) relative to its 
southern counterpart ($\alpha=-0.88$). (Both of these spectral indices are 
taken from PNSS.) It therefore appears as if the northern lobe is not at 
present being fuelled. However, this notion is at odds with the orientation 
of the jet, which points to that lobe. The simplest, but not necessarily 
correct, way to overcome this difficulty is to assume that the jet and the 
lobe on its side are aggregated and that they both form an FR\,I-type 
structure. Under this assumption, J1211+743 should consequently be labelled 
a HYbrid MOrphology Radio Source (HYMORS), i.e. a source each of whose sides 
are of a different FR type \citep{GKW2000,Gawr2006}.

The hypothesis that J1211+743 is a HYMORS is, however, not entirely 
justified. Firstly, the configuration of magnetic fields along the jet is 
typical of those in FR\,II-type sources \citep{vBW1981}. Therefore, one has 
to make a speculative assumption that the jet in J1211+743 is highly 
dissipative for the lobe facing it to be diffuse. Another problem -- already 
noted by PNSS based on their measurements -- is that the values of the 
low-frequency spectral indices at the outer edges of the lobes are 
consistent with those provided by low-frequency observations of luminous 
FR\,II radio sources. These circumstances are clearly incompatible with the 
assumption that the northern lobe of J1211+743 is of FR\,I type. This 
conclusion is additionally supported by the low-resolution WSRT maps of 
\citet{vBW1981}, particularly their 609-MHz map, where the source appears as 
a triple that does not resemble an FR\,I radio galaxy. It follows that there 
are insufficient grounds for aggregation of the jet and the lobe on its 
side. These two components seem to be separate entities and their 
connection, likely caused by the projection, is only apparent. On the basis 
of the appearance of the source in the 607-MHz GMRT map by PNSS, it could be 
even argued that there is no physical connection at all, i.e. the tip of the 
jet is located outside the boundaries of the northern lobe. If this is the 
case, the jet is not energising the lobe and the alignment of these two 
components is merely coincidental.

As J1211+743 does not appear to be a HYMORS, in Sect.\,2 I~present an 
alternative interpretation of its radio morphology that consistently 
explains the nature of this source.

\section{A model based on an activity restart}

\subsection{The concept}

Assume that the activity of the central engine of J1211+743 ceased at some 
point in the past. When this happened, i.e. when the nucleus ended its 
supply of relativistic plasma via jets to the hotspots, the lobes started 
to become weak and diffuse, while their spectra steepened owing to 
radiation and expansion losses. The lobes are, however, huge reservoirs of 
energy, so even when not being fed by the jets, they are observable for a 
substantial amount of time, up to $10^8$\,yr according to \citet{KG1994}. 
The hotspots fade in radio brightness much sooner than the lobes -- their 
lifetimes are roughly $7\times 10^4$\,yr \citep{KSR2000}, hence they are 
normally not seen in coasting lobes. Therefore, a faint, diffuse, and 
steep-spectrum relic lobe devoid of a hotspot is a good signature of the 
energy cut-off \citep{KC2002}.

If a coasting, large-scale, double-lobed radio source does not lie close to 
the sky plane, then the light-travel time difference between the images of the 
lobes is significant: they are observed at two completely different 
stages of their decay. For typical linear sizes of a large-scale FR\,II 
source (hundreds of kpc) and assuming even quite moderate angles between 
the source axis and the sky plane, the time lag may
easily become longer than the aforementioned lifetime of a hotspot. Thus, the 
far-side lobe may still be perceived as a typical FR\,II-like with a 
hotspot, whereas the near-side one, representing a much later stage of the 
evolution of the radio source, may be seen as a diffuse relic without a 
hotspot. This argument was already used by \citet{MS11} to explain the
lobe asymmetry in radio sources they studied.

On the basis of the analysis of their data, PNSS conclude that the axis of 
J1211+743 makes an angle close to 45$\degr$ with the line of sight. It follows
that the difference between the radial distances to the extreme regions
(``tips'') of the lobes is roughly equal to the projected linear size of the
source. Because of this size (846\,kpc), we perceive radiation from the tip
of the far-side lobe as being delayed by $2.76\times10^6$ yr relative to 
that from the tip of the near-side lobe. The length of this period exceeds 
hotspot longevity by 1.6\,dex, so it is quite likely that the observer finds 
that the near-side hotspot disappears while the far-side hotspot remains in 
place, which is exactly what is observed in J1211+743. On the other hand, 
the $2.76\times10^6$-year lag between the images of the lobes falls short by 
$\sim 1.5$\,dex of the time required for the near-side lobe to disperse 
fully \citep{KG1994}, so given that the far-side lobe resembles a standard 
FR\,II, the near-side one must be visible too, which is what we see. Hence, 
the lobe asymmetry is convincingly explained by activity cessation.

The activity can be restarted, however, leading to the emergence of a newly 
formed jet. Since the orientation of the source with regard to 
the line of sight is close to 45$\degr$ (PNSS), i.e. close to the 
galaxy/quasar dividing line \citep{Bart1989}, the jet is one-sided and 
indicates which side of the radio source is closest to the observer. 
Consequently, the following spatial coincidence occurs: the jet and the 
decaying lobe lie on the same side. This coincidence is an analogue of the 
Laing-Garrington effect \citep{Laing1988,Garr1988}, but here, instead of the 
correlation being between the jet orientation and polarisation asymmetry, 
there is a correlation between the jet orientation and morphological asymmetry.

I~now present a detailed timeline of the evolution of J1211+743, which is 
illustrated with a schematic view of the source (Fig.\,1). This diagram 
is by no means a paraphrasing of the radio map of J1211+743. While the map 
is a projection of the source onto the sky plane, the plane of the drawing 
in Fig.\,1 is defined so as to include the observer, the core of J1211+743, 
the line of sight, both lobes, and the axis connecting them\footnote{The 
projections of the lobes in the sky plane also lie in the plane of the 
drawing in Fig.\,1 but are not shown for clarity's sake.}. In other words, 
the normal to the plane of the drawing in Fig.\,1 is tangential to the 
celestial sphere at the position of J1211+743 and perpendicular to the 
source's projected axis. The north-western (near-side) lobe is at the lower
right of Fig.\,1, while the south-eastern (far-side) lobe is at the upper 
left.

\subsection{The timeline}

\begin{enumerate}

\item The nucleus of J1211+743 was originally active, producing jets for
a sufficiently long period of time to inflate the large lobes we observe now. 
Their separation is also very large: assuming an angle of 45$\degr$ between the 
source axis and the line of sight (PNSS) and given the projected linear size 
of 846\,kpc, the actual linear size of the whole radio structure is 1.2\,Mpc. 
Hence, J1211+743 could be labelled a giant radio galaxy.

\item The activity of the nucleus ceased at some point in the past, which led
to the cut-off of the stream of relativistic plasma. The information about 
this event is carried by the last jet particles to be emitted moving at the 
velocity $\beta_{jet} c$ where $\beta_{jet} \lessapprox 1$, i.e. it 
travelling through the jets' channels towards the lobes with a velocity 
close to the speed of light. Empty jet channels collapse quickly -- the 
timescale of this process is of the order of 10$^4$ yr \citep{Brock2011}.

\item After time period $t_1=l_1/\beta_{jet} c$, the information that there 
is no more energy supply reaches the hotspot of the far-side lobe. 
Similarly, after $t_2=l_2/\beta_{jet} c$, analogous information arrives at 
the hotspot of the near-side lobe. The radio emission carrying the 
information about the status (images) of the far-side and near-side lobes at 
the instants when each of them has just begun to no longer be fuelled is 
represented by wavefronts ``1'' and ``3'', respectively (see Fig.\,1). As 
can be clearly seen in Fig.\,1, $l_1 < l_2$, which is consistent with the 
images of J1211+743, particularly those in PNSS. For simplicity, I~assume 
that the arm-length asymmetry is entirely intrinsic, and thus that wavefront
``1'' departs its lobe prior to wavefront ``3''.

\item After $t_{decay}=7\times 10^4$\,years relative to the events 
represented by wavefronts ``1'' and ``3'', the hotspots faded in radio 
brightness. The radio emission carrying the information about each of these 
two events is represented by wavefronts ``2'' and ``4'', respectively.

\item Meanwhile, the activity of the nucleus (denoted as ``core'' in 
Fig.\,1) had restarted. I~assume that the period of quiescence of the nucleus 
$t_q$ was longer that the lifetime of the jet channels (10$^4$\,yr), so they 
had collapsed before the activity re-ignition and the new jets had to 
``drill'' them anew. Nevertheless, the advance velocity of the jet, 
$\beta_{adv}c$ in such circumstances is considerably larger than 0.03\,$c$, 
which is typical of lobes advancing through an intact intergalactic medium. 
For example, in a well-known double-double radio galaxy B\,1834+620, this 
velocity is in the range $0.19 < \beta_{adv} < 0.29$ \citep{Sch2000}.

\item At a certain instant, the length of the jet attains $d$ and
radiation carrying the information about the location of its tip is 
emitted. This event is represented by wavefront ``5'', which is the one 
currently passing the observer.

\end{enumerate}

\begin{figure*}
\vspace{-1cm}
\centering
\includegraphics[width=1.44\columnwidth]{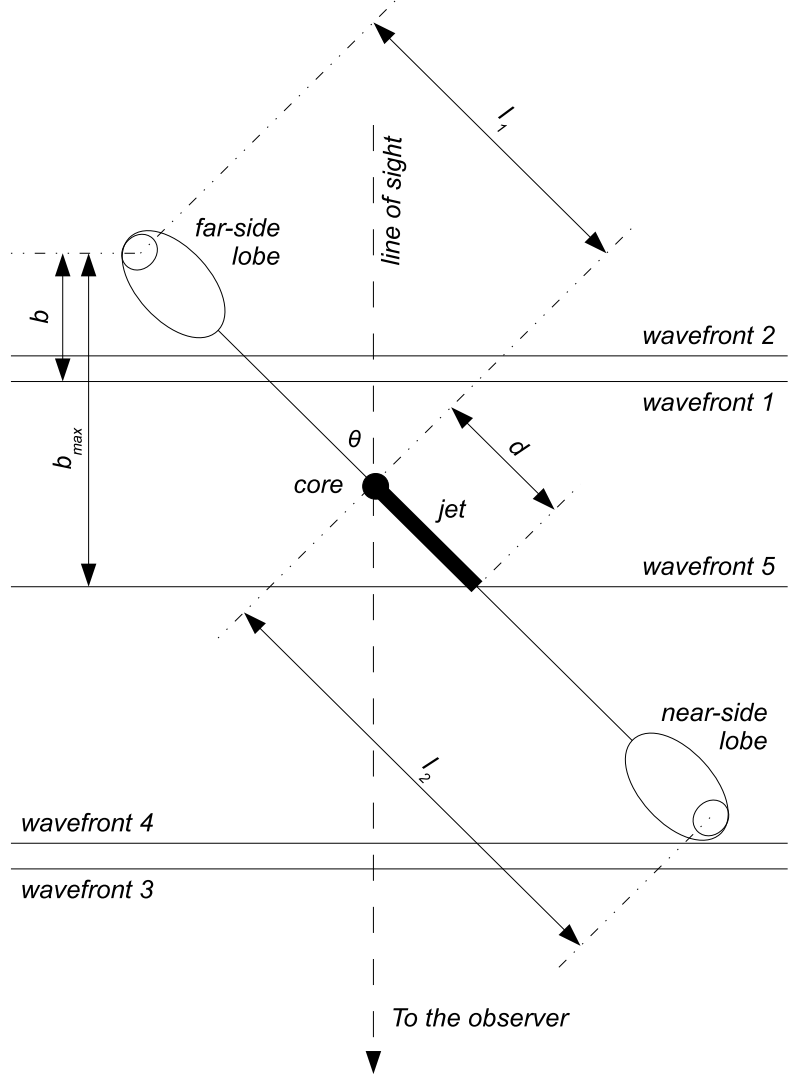}
\vspace{-2.5cm}
\caption{A schematic diagram of J1211+743. (Lengths and distances are not
drawn to scale.)}
\end{figure*}

Our model is compatible with the observations only if wavefront ``5'' lies 
between wavefronts ``1'' and ``4'', as shown in Fig.\,1. In other words, as 
wavefront ``5'' passes the observer, wavefront ``4'' must have already 
passed, i.e. the near-side hotspot appears to be already dispersed, whereas 
wavefront ``1'' has not yet passed by, i.e. the far-side hotspot remains 
visible. To show that such a situation is possible, a quantitative analysis 
of our model must be carried out.

\subsection{Quantitative analysis}

Before proceeding with the quantitative analysis, I~note one assumption that 
is made throughout. As can be seen in Fig.\,1, the axis of the source and 
the jet make the same angle $\theta$ with the line of sight. Whether this is 
indeed the case for J1211+743 cannot be easily verified so this is only an 
assumption. In making it, I~rely on the presumed analogy between J1211+743 
and double-double sources, where activity restart surely takes place 
\citep[see the review of][]{SJ2009} and where a good alignment between the 
axis of an outer (old) and an inner (new) pair of lobes is observed. 
However, it cannot be ruled out that the jet in J1211+743 is ejected in a 
different direction and that the ``old'' and the ``new'' axes coincide only 
apparently owing to their particular orientations and projections on the sky 
plane that conceal this difference. Moreover, each arm of the radio source may
in principle have a different value of $\theta$, but since I~carried out a
separate analysis for each lobe, this did not introduce uncertainty into my
results.

As a first step of the quantitative analysis, I~verify whether wavefront 
``4'' always precedes wavefront ``5''. This condition is fulfilled if the 
time of propagation of the information about the energy cutoff in the 
channel of the near-side jet $t_2$ when increased by the time of hotspot 
dispersion $t_{decay}$ does not exceed the sum of the durations of the 
quiescence period $t_q$, the time needed by the jet to attain its present 
length $d$, and the time wavefront ``5'' spends travelling before it reaches 
the point along the line of sight that wavefront ``4'' started from, given 
by

\begin{equation}
t_2 + t_{decay} \leq t_q + {d\over\beta_{adv} c} + {(l_2 - d)\cos\theta\over c} .
\end{equation}
Solving Eq.\,(1) for $t_q$, one gets

\begin{equation}
t_q \geq {1\over c} \big[{l_2\over\beta_{jet}} + c t_{decay} -
{d\over\beta_{adv}} - (l_2 - d)\cos\theta \big].
\end{equation}
From the observations, we have $l_2\sin\theta=484$\,kpc (PNSS) and 
$d\sin\theta=160$\,kpc \citep{vBW1981}. Using these values for $l_2$ and 
$d$, and substituting any value of $\beta_{adv}$ from the range obtained by 
\citet{Sch2000} and $\beta_{jet}$ even as low as 0.9, I~found that if only 
$\theta < 88.7\degr$ then Eq.\,(2) was fulfilled even for the lowest 
possible $t_q=10^4$\,yr. However, given the lack of a counter-jet in 
J1211+743, it is highly unlikely that it lies close to the sky plane, and so 
it is almost certain that $\theta$ is far below $88.7\degr$. It follows that 
Eq.\,(2) is always fulfilled.

Now I~investigate the necessary conditions for wavefront ``1'' remain behind 
wavefront ``5'': $b \leq b_{max}$, where $b_{max}=(l_1+d)\cos\theta$.
On a purely qualitative basis, this implies that there is 
an upper limit to $t_q$, because if the activity is restarted after an 
excessively long break, and so the jet starts its build-up too late, it will 
be unable to attain the observed length $d$ before the arrival of the 
information that the far-side hotspot has started to fade out. Therefore,
I~will calculate the upper limit to $t_q$ and check whether it is longer than 
the lower limit $t_q=10^4$\,yr. If it appears to be the case, i.e. if a 
range of allowable values of $t_q$ does exist, we will have proof that our 
scenario is plausible.

The condition $b \leq b_{max}$ is fulfilled if the sum of the duration 
of quiescence period $t_q$ and the time needed by the jet to attain its 
present length $d$ does not exceed the sum of the time of propagation of 
the information about the energy cutoff in the channel of the far-side jet 
$t_1$ and the light-travel time over the distance $b_{max}$

\begin{equation}
t_q + {d\over\beta_{adv} c} \leq t_1 + {b_{max}\over c}.
\end{equation}
Solving Eq.\,(3) for $t_q$, one gets

\begin{equation}
t_q  \leq {1\over c} \big[{l_1\over\beta_{jet}} - {d\over\beta_{adv}} + (l_1 
+ d)\cos\theta \big].
\end{equation}
When one substitutes $l_1$=512\,kpc and $d$=226\,kpc\footnote{Both these 
figures have been corrected for $1/\sin\theta$, where $\theta=45\degr$.} 
from the observations by PNSS and \citet{vBW1981}, respectively, into 
Eq.\,(4), one finds that this relation is fulfilled for $\beta_{adv} \geq 
0.219$. This is close to the smallest value of $\beta_{adv}$ given by 
\citet{Sch2000} for B\,1834+620. On the other hand, $\beta_{adv}=0.29$, i.e. 
the largest value of $\beta_{adv}$ according to \citet{Sch2000}, yields the 
upper limit $t_q=8.3\times 10^5$\,yr. This proves that our scenario is 
plausible.

The above result has been achieved under the assumption made by PNSS that 
$\theta=45\degr$. Additionally, $\beta_{jet}=1$ has been assumed for 
simplicity. I~now test whether our model remains plausible if three 
parameters in Eq.\,(4) -- $\beta_{jet}$, $\beta_{adv}$, and $\theta$ -- are 
varied. To this end, I~substitute values from the range $10\degr\leq 
\theta\leq 80\degr$ and $0.2 \leq \beta_{adv} \leq 0.35$ into Eq.\,(4), and 
calculate the upper limits to $t_q$ (in Myr) for $\beta_{jet}=1$ and 
$\beta_{jet}=0.9$. The results for different combinations of $\theta$, 
$\beta_{adv}$, and $\beta_{jet}$ are given in Table\,1. The lack of a number 
at the crossing of a given row and column means that the respective 
combination of parameters is not allowed if the scenario is to be viable.

\section{Conclusion}

Recurrent activity in active galactic nuclei is a well-known phenomenon 
\citep{SJ2009}. In this letter, I~have demonstrated that the radio 
morphology of J1211+743 can be understood in the framework of a restarted 
activity scenario combined with the assumption that the radio source does 
not lie in the sky plane. Owing to the latter circumstance, each lobe is 
observed at a different epoch and this renders them a different appearance. 
On the other hand, the activity re-ignition is manifested by the presence of 
the jet whose direction, apparently paradoxically, correlates with the 
location of the relic lobe.

Our model requires that there is an upper limit to duration of the quiescent 
period between the two active periods. This limit strongly depends on the 
jet advance velocity $\beta_{adv}$ and inclination angle $\theta$. As can be 
seen in Table\,1, for a wide range of values of $\theta$ and for
several values of $\beta_{adv}$ typical of restarted sources known from the 
literature, the upper limits are of the order of about $10^6$ years. This 
means that our model is realistic, because quiescent periods of this length 
are allowed. Depending on the mechanism, activity intermittency can operate 
on very different timescales that are either long in the case of ionisation 
instability in the accretion disk \citep[][and references therein]{HSE2001, 
Janiuk04} or short in the case of radiation-pressure instability 
\citep{Czerny09}.

\begin{table}[t]
\caption{Upper limits to the length of the quiescent period [Myr]}
\begin{tabular}{c|c c c c|c c c c}
\hline
\hline
         & \multicolumn{4}{c|}{$\beta_{jet}=1$}
         & \multicolumn{4}{c}{$\beta_{jet}=0.9$}\\
           \cline{2-9}
$\theta$ & \multicolumn{4}{c|}{$\beta_{adv}$}
         & \multicolumn{4}{c}{$\beta_{adv}$}\\
$[\degr]$ & 0.2 & 0.25 & 0.3 & 0.35 & 0.2 & 0.25 & 0.3 & 0.35 \\
\hline
10 & 1.43 & 4.43 & 6.43 & 7.86 & 2.18 & 5.19 & 7.19 & 8.62 \\
20 & 0.50 & 2.03 & 3.04 & 3.77 & 0.88 & 2.41 & 3.43 & 4.15 \\
30 & 0.09 & 1.13 & 1.83 & 2.33 & 0.35 & 1.40 & 2.09 & 2.59 \\
40 & ---  & 0.62 & 1.16 & 1.55 & 0.01 & 0.82 & 1.36 & 1.75 \\
50 & ---  & 0.24 & 0.70 & 1.02 & ---  & 0.42 & 0.87 & 1.19 \\
60 & ---  & ---  & 0.34 & 0.62 & ---  & 0.09 & 0.49 & 0.78 \\
70 & ---  & ---  & 0.02 & 0.29 & ---  & ---  & 0.16 & 0.43 \\
80 & ---  & ---  & ---  & ---  & ---  & ---  & ---  & 0.12 \\
\hline
\hline
\end{tabular}
\end{table}

\begin{acknowledgements}

This research has made use of the NASA/IPAC Extragalactic Database (NED) 
which is operated by the Jet Propulsion Laboratory, California Institute of 
Technology, under contract with the National Aeronautics and Space 
Administration.

\end{acknowledgements}

\end{document}